# Zero-Shot Whole Slide Image Retrieval in Histopathology Using Embeddings of Foundation Models

Saghir Alfasly[1], Ghazal Alabtah[1], Sobhan Hemati[1], Krishna Rani Kalari[2], H.R. Tizhoosh[1]

[1] Kimia Lab, Dept. of Artificial Intelligence & Informatics
Mayo Clinic, Rochester, MN, USA

[2] Division of Computational Biology, Dept. of Quantitative Health Sciences
Mayo Clinic, Rochester, MN, USA

# Abstract

We have tested recently published foundation models for histopathology for image retrieval. We report macro average of F1 score for top-1 retrieval, majority of top-3 retrievals, and majority of top-5 retrievals. We perform zero-shot retrievals, i.e., we do not alter embeddings and we do not train any classifier. As test data, we used diagnostic slides of TCGA, The Cancer Genome Atlas, consisting of 23 organs and 117 cancer subtypes. As a search platform we used Yottixel that enabled us to perform WSI search using patches. Achieved F1 scores show low performance, e.g., for top-5 retrievals, 27% ± 13% (Yottixel-DenseNet), 42% ± 14% (Yottixel-UNI), 40%±13% (Yottixel-Virchow), 41%±13% (Yottixel-GigaPath), and 41%±14% (GigaPath WSI).

# Introduction

The emergence of digital pathology and artificial intelligence (AI) seems to bring about an eminent revolution in the field of histopathology, with anticipated efficiency and accuracy at unprecedented levels for diagnostic and treatment planning of many diseases. Digital pathology offers high-resolution tissue images that can be processed and analyzed by advanced computer algorithms. This shift could enable pathologists to leverage the power of deep learning models to identify tissue patterns, classify diseases, and predict patient outcomes with greater precision. In this context, image retrieval may also play a critical role in enhancing diagnostic workflows, provided we can finally close the semantic gap after three decades of research.

Image search and retrieval in histopathology is particularly significant as it allows pathologists to search databases - small or large - of digital slides to find visually, i.e., anatomically, similar cases. This capability is invaluable for comparative diagnosis, where finding similar cases can aid in confirming or refining a diagnosis. However, the effectiveness of image retrieval depends, among others, on the quality of embeddings produced by deep neural networks. High-quality embeddings ensure that similar images are clustered closely together in the feature space, leading to more accurate retrieval results. Conversely, poor-quality embeddings can result in misclassification and retrieval of irrelevant cases, potentially compromising diagnostic accuracy.

The advent of foundation models, large-scale AI models pre-trained on vast amounts of data, holds great promise in addressing the long-standing semantic gap problem in image search and retrieval for histopathology [Bommasani2021]. The semantic gap refers to the disparity between the high-level concepts understood by human experts and the low-level features captured by machines. Foundation models, with their expected ability to generalize across diverse tasks and domains, could bridge this gap by learning more robust and semantically meaningful representations of medical images. This development offers hope that AI-driven image retrieval will become more accurate and reliable, ultimately leading to better patient outcomes and more efficient healthcare systems.

## Results

Table 1 shows the overall results for all WSI retrievals. Total average and standard deviation, and 95% confidence intervals of F1 scores across all organs and subtypes for WSI retrievals for all Yottixel variations and GigaPath are reported.

*Table 1. The macro average and 95% confidence internal of F1 scores for tested WSI retrievals. Yottixel, in its default configuration using pre-trained DenseNet, serves as the testbed and benchmark. DenseNet was replaced by UNI, Virchow and GigaPath to provide Yottixel with patch embeddings. WSI retrievals were performed using Yottixel's median-of-minimums scheme. GigaPath WSI (last row) is the only results independent from Yottixel as GigaPath can provide one embedding for WSI. The results for GigaPath WSI are reported for 17 organs out of 23 due to significant computational resources required for processing. Results will be updated upon availability if all experimental data.*

|  | Top-1 | Majority @ Top-3 | Majority @ Top-5 |
|---|---|---|---|
| Yottixel | 28% ± 13% [23% 33%] | 28% ± 13% [17% 32%] | 27% ± 13% [22% 32%] |
| Yottixel-UNI | 44% ± 15% [38% 50%] | 44% ± 14% [38% 50%] | 42% ± 14% [36% 48%] |
| Yottixel-Virchow | 41% ± 15% [35% 47%] | 41% ± 14% [35% 47%] | 40% ± 13% [36% 45%] |
| Yottixel-GigaPath | 43% ± 15% [37% 49%] | 43% ± 15% [37% 49%] | 41% ± 13% [35% 47%] |
| GigaPath WSI* | 43% ± 16% [37% 50%] | 42% ± 16% [36% 48%] | 41% ± 14% [35% 47%] |

\* GigaPath WSI run for 17 organs only; remaining results to follow.

Table 2, Table 3 and Table 4 show the detailed results for top-1, majority of top-3 and majority of top-5 accuracy, respectively. All models improve the results compared to DenseNet (default configuration of Yottixel). As DenseNet is a much smaller network trained with ImageNet natural images, these improvements are expected. However, with the exception of kidney, all values are quite low.

Table 2. Macro average of F1 score for top-1 WSI retrievals. Best results highlighted for each organ. The results for GigaPath WSI are reported for 17 organs out of 23 due to significant computational resources required for processing. Results will be updated upon availability if all experimental data.

| organ | Yottixel | Yottixel+UNI | Yottixel+Virchow | Yottixel+GigaPath | GigaPath-WSI |
|---|---|---|---|---|---|
| Adrenal glands | 0.24 | 0.42 | 0.37 | 0.40 | 0.38 |
| Bladder | 0.53 | 0.66 | 0.65 | 0.63 | 0.67 |
| Brain | 0.26 | 0.39 | 0.44 | 0.42 | - |
| Breast | 0.17 | 0.27 | 0.25 | 0.26 | 0.25 |
| Cervix | 0.15 | 0.27 | 0.27 | 0.31 | 0.31 |
| Colon | 0.54 | 0.61 | 0.58 | 0.60 | 0.54 |
| Esophagus | 0.41 | 0.63 | 0.61 | 0.78 | 0.71 |
| Eye | 0.20 | 0.41 | 0.37 | 0.42 | 0.34 |
| Head and neck | 0.23 | 0.44 | 0.37 | 0.46 | - |
| Kidneys | 0.50 | 0.82 | 0.78 | 0.71 | 0.73 |
| Lining | 0.37 | 0.35 | 0.32 | 0.31 | 0.30 |
| Liver | 0.33 | 0.33 | 0.33 | 0.33 | 0.33 |
| Lungs | 0.12 | 0.21 | 0.21 | 0.21 | - |
| Pancreas | 0.26 | 0.50 | 0.47 | 0.50 | 0.5 |
| Prostate | 0.50 | 0.50 | 0.50 | 0.50 | 0.5 |
| Rectum | 0.26 | 0.66 | 0.55 | 0.63 | 0.57 |
| Skin | 0.20 | 0.30 | 0.22 | 0.26 | 0.24 |
| Soft tissues | 0.18 | 0.43 | 0.37 | 0.42 | - |
| Stomach | 0.14 | 0.30 | 0.29 | 0.30 | 0.34 |
| Testicles | 0.23 | 0.35 | 0.36 | 0.35 | 0.34 |
| Thymus | 0.11 | 0.29 | 0.25 | 0.26 | 0.27 |
| Thyroid gland | 0.32 | 0.47 | 0.41 | 0.42 | - |
| Uterus | 0.26 | 0.52 | 0.45 | 0.49 | - |

Table 3. Macro average of F1 score for majority of top-3 WSI retrievals. Best results highlighted for each organ. The results for GigaPath WSI are reported for 17 organs out of 23 due to significant computational resources required for processing. Results will be updated upon availability if all experimental data.

| organ | Yottixel | Yottixel+UNI | Yottixel+Virchow | Yottixel+GigaPath | GigaPath-WSI |
|---|---|---|---|---|---|
| Adrenal glands | 0.25 | 0.43 | 0.38 | 0.41 | 0.31 |
| Bladder | 0.50 | 0.68 | 0.66 | 0.66 | 0.66 |
| Brain | 0.28 | 0.43 | 0.44 | 0.42 | - |
| Breast | 0.16 | 0.30 | 0.23 | 0.25 | 0.25 |
| Cervix | 0.12 | 0.30 | 0.30 | 0.35 | 0.33 |
| Colon | 0.50 | 0.58 | 0.57 | 0.58 | 0.51 |
| Esophagus | 0.37 | 0.61 | 0.60 | 0.76 | 0.69 |
| Eye | 0.20 | 0.37 | 0.39 | 0.43 | 0.42 |
| Head and neck | 0.23 | 0.43 | 0.30 | 0.43 | - |
| Kidneys | 0.50 | 0.78 | 0.77 | 0.70 | 0.73 |
| Lining | 0.38 | 0.35 | 0.32 | 0.37 | 0.26 |
| Liver | 0.33 | 0.33 | 0.33 | 0.33 | 0.33 |
| Lungs | 0.11 | 0.22 | 0.23 | 0.22 | - |
| Pancreas | 0.22 | 0.47 | 0.50 | 0.50 | 0.48 |
| Prostate | 0.50 | 0.5 | 0.50 | 0.50 | 0.50 |
| Rectum | 0.28 | 0.57 | 0.52 | 0.53 | 0.50 |
| Skin | 0.21 | 0.30 | 0.23 | 0.24 | 0.21 |
| Soft tissues | 0.18 | 0.45 | 0.38 | 0.42 | - |
| Stomach | 0.15 | 0.35 | 0.32 | 0.30 | 0.32 |
| Testicles | 0.26 | 0.38 | 0.37 | 0.34 | 0.37 |
| Thymus | 0.11 | 0.32 | 0.29 | 0.28 | 0.31 |
| Thyroid gland | 0.32 | 0.47 | 0.42 | 0.43 | - |
| Uterus | 0.26 | 0.52 | 0.44 | 0.52 | - |

Table 4. Macro average of F1 score for majority of top-5 WSI retrievals. Best results highlighted for each organ. The results for GigaPath WSI are reported for 17 organs out of 23 due to significant computational resources required for processing. Results will be updated upon availability if all experimental data.

| organ | Yottixel | Yottixel+UNI | Yottixel+Virchow | Yottixel+GigaPath | GigaPath-WSI |
|---|---|---|---|---|---|
| Adrenal glands | 0.25 | 0.36 | 0.35 | 0.34 | 0.29 |
| Bladder | 0.48 | 0.70 | 0.63 | 0.69 | 0.65 |
| Brain | 0.29 | 0.44 | 0.43 | 0.45 | - |
| Breast | 0.15 | 0.31 | 0.24 | 0.26 | 0.27 |
| Cervix | 0.11 | 0.23 | 0.29 | 0.28 | 0.30 |
| Colon | 0.46 | 0.56 | 0.54 | 0.55 | 0.56 |
| Esophagus | 0.42 | 0.61 | 0.59 | 0.61 | 0.58 |
| Eye | 0.27 | 0.36 | 0.36 | 0.35 | 0.36 |
| Head and neck | 0.23 | 0.28 | 0.29 | 0.39 | - |
| Kidneys | 0.52 | 0.76 | 0.68 | 0.67 | 0.71 |
| Lining | 0.27 | 0.38 | 0.37 | 0.37 | 0.30 |
| Liver | 0.33 | 0.33 | 0.33 | 0.33 | 0.33 |
| Lungs | 0.11 | 0.20 | 0.21 | 0.18 | - |
| Pancreas | 0.23 | 0.47 | 0.45 | 0.45 | 0.40 |
| Prostate | 0.50 | 0.50 | 0.50 | 0.50 | 0.50 |
| Rectum | 0.26 | 0.56 | 0.37 | 0.53 | 0.46 |
| Skin | 0.19 | 0.23 | 0.23 | 0.24 | 0.19 |
| Soft tissues | 0.19 | 0.46 | 0.37 | 0.40 | - |
| Stomach | 0.14 | 0.35 | 0.29 | 0.30 | 0.33 |
| Testicles | 0.25 | 0.41 | 0.39 | 0.36 | 0.37 |
| Thymus | 0.10 | 0.36 | 0.32 | 0.29 | 0.37 |
| Thyroid gland | 0.33 | 0.44 | 0.44 | 0.43 | - |
| Uterus | 0.21 | 0.46 | 0.43 | 0.45 | - |

# Methods

**Evaluation** – We used macro average of F1-score for top-1, majority of top-3, and majority of top-5 WSI retrievals. As the data is imbalanced, using macro average is a rigorous evaluation.

**Search Engine** - To evaluate the embeddings of models in a zero-shot manner, we needed to choose a dependable search platform that met several essential criteria. First, the platform should allow the integration of a new model into its structure without disrupting its overall design and requiring minimal additional ablation studies. Second, it should have an unsupervised patching algorithm. Third, the platform should be capable of performing patch searches. Fourth, it should support whole-slide image (WSI) searches by utilizing all selected patches. Lastly, the platform should have a proven track record of storage efficiency and search speed, and overall higher accuracy compared to other search engines. After considering these factors, we selected Yottixel for our testing [Kalra2020a, Tizhoosh2024, Lahr2024]. As Figure 1 illustrates, Yottixel topology can easily integrate any deep model.

**Data** – We used The Cancer Genome Atlas diagnostic slides consisting of 11,444 WSIs of 9,339 patients covering 23 organs and 117 cancer subtypes. Tables TA1 to TA23 in the Appendix provide details of all subtypes for each organ, number of patients and whole slide images. TCGA Projects (e.g., lung adenocarcinoma, LUAD, and lung squamous cell carcinoma, LUSC) are very generic terms, using large buckets for classification. For example, in lung Adenocarcinoma and Squamous Cell Carcinoma cover about 98% of cases; however, the subtypes are incredibly helpful to understand biology and prognosis—and ultimately treatment. For instance, bronchio-alveolar adenocarcinoma (now called *lepidic*) is typically treated successfully with surgery alone. Adenocarcinoma with mixed subtypes is more likely to be larger, present with metastases, and show worsened clinical outcomes. Most studies have rather used the generic tumor types and not detailed subtyping for retrieval (e.g., see [Kalra2020b]).

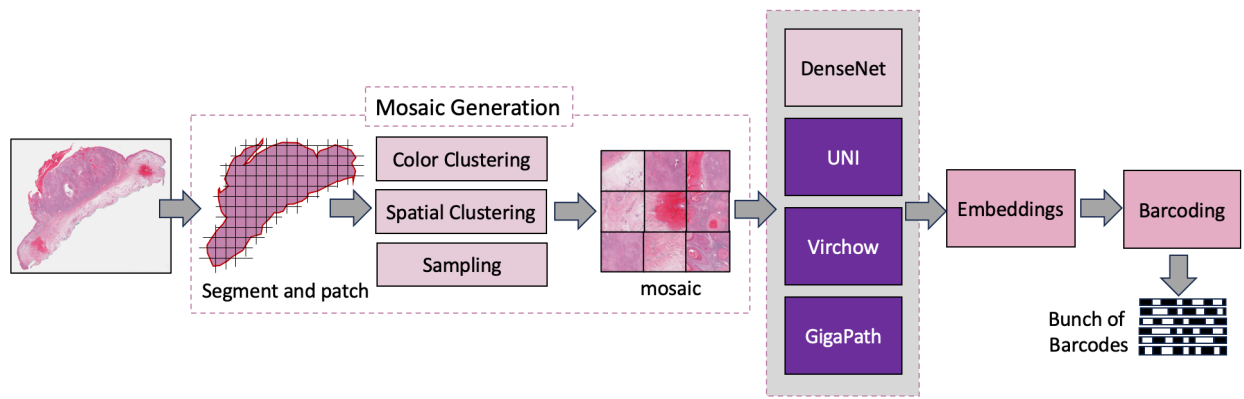

*Figure 1. Yottixel indexing and search allows easy integration of any deep model. The "mosaic" generation provides a small set of representative patches that can be fed into any deep model (DenseNet in original Yottixel configuration). Foundation models such as UNI, Virchow and GigaPath can simply be integrated with Yottixel structure without inferring with its overall design. Barcoding of patch embeddings will deliver "bunch of barcodes" as the index of WSI. Through "medium-of-minimum" Hamming distance calculation, WSIs can be compared.*

**Patching WSIs** – We used the Yottixel patching method, which consists of two unsupervised stages. In the first stage, we segment the whole slide image (WSI) into different regions based on color composition using k-means clustering. We do this to capture the pattern variability within the WSI from a computer vision perspective, separating regions like blood stains, muscle tissue, fat, etc. Based on my visual inspection of different WSIs, we typically set the number of color clusters to 9 [Kalra2020a]. In the second stage, we select a small percentage of representative patches from each color-segmented region while preserving spatial diversity. We accomplish this using k-means clustering again, but this time we group patches based on their location within each color region. Unlike the default settings in the Yottixel paper that uses 1024x1024 patches, we output patches in 224x224 pixel dimensions to accommodate the foundation models, and we use 2% of the representative patches instead of 5%. The result is a "mosaic" of patches that represent the entire WSI in a compact manner. This mosaic dramatically reduces the computational burden compared to processing the entire gigapixel WSI, while still capturing the key visual patterns and structures present in the slide.

**Models for Search Engine**

**UNI** is a self-supervised vision encoder based on a large Vision Transformer (ViT-L) architecture [Chen2024]. It was pretrained on the Mass-100K dataset, which comprises over 100 million tissue patches extracted from 100,426 H&E whole-slide images (WSIs) spanning 20 major tissue types. The pretraining process utilized the DINOv2 self-supervised learning approach, which employs student-teacher knowledge distillation and masked image modeling. UNI's training involved two main loss objectives: self-distillation loss and masked image modeling loss. The model was developed using a total of 100,130,900 images at various resolutions, including 75,832,905 images at 256x256 pixels and 24,297,995 images at 512x512 pixels, all at 20x magnification.

**Virchow** is a large-scale self-supervised vision model for computational pathology [Vorontsov2024]. It is based on a Vision Transformer (ViT) architecture, specifically using the ViT-H/14 (ViT-Huge) configuration with 632 million parameters. Virchow was pretrained on a dataset of approximately 1.5 million H&E stained whole-slide images (WSIs) from about 100,000 patients, spanning 17 high-level tissue types. The pretraining data includes both cancerous and benign tissues, collected via biopsy (63%) and resection (37%). Similar to UNI, the model was trained using the DINO v.2 self-supervised learning framework. Virchow processes 224x224 pixel tissue tiles extracted from WSIs at 20x magnification (0.5 microns per pixel). The training involved approximately 2 billion tiles sampled from a pool of about 13 billion available tissue tiles.

**GigaPath** is a vision transformer architecture for pretraining large pathology foundation models on gigapixel whole-slide images (WSIs) [Xu2024]. It is pretrained on an extensive dataset comprising 1.3 billion 256x256 image tiles extracted from 171,189 pathology slides sourced from over 30,000 patients across 28 cancer centers in the Providence health network. This dataset encompasses 31 major tissue types, significantly surpassing the scale of previous datasets like TCGA. GigaPath employs a two-stage pretraining approach, beginning with image-level self-supervised learning using DINOv2 and progressing to whole-slide-level self-supervised learning through a masked autoencoder combined with the LongNet architecture. This adaptation allows for effective ultra-large-context modeling, capturing both local and global patterns across the entire slide. GigaPath's architecture enables it to process tens of thousands of image tiles, resulting in state-of-the-art performance across various digital pathology tasks.

Table 5. Details of models used by Yottixel to acquire patch embedding. GigaPath-WSI (the last row) is independent from Yottixel as it can generate a single embedding for WSI.

|  | Size | GPU Time | Embed Size | Storage (BoB) |
|---:|---|---|---|---|
| DensNet | 6,953,856 | 30 minutes | 1024 | 1500MB |
| UNI | 303,350,784 | 552 minutes | 1024 | 1500MB |
| Virchow | 631,229,184 | 1,247 minutes | 1280 | 1983MB |
| GigaPath | 1,134,953,984 | 1,589 minutes | 1536 | 2382MB |
| GigaPath-WSI | 1,221,284,864 | - | 786 | 1.65MB |

# Discussions

Accurately subtyping cancerous tissue remains a significant challenge in diagnostic pathology. The potential of image retrieval techniques as tools for aiding diagnosis, through similarity matching with evidently diagnosed past cases, has garnered attention in recent years. Foundation models trained on vast collections of whole slide images (WSIs), expected to deliver expressive and semantically correct embeddings, have generated optimism about the feasibility of clinical-grade image retrieval systems. However, findings from this validation study reveal that there is still considerable progress to be made before these models can be effectively utilized in clinical settings. The study reported an average F1 score of approximately 44% for zero-shot WSI retrieval tasks, indicating that the current performance of these models is not yet adequate for clinical application.

It is important to contextualize these findings by considering the design and intended use of the foundation models, such as UNI, Virchow, and patch-based backbone of GigaPath, which were primarily developed for patch-level analysis and representation learning, not for WSI-level retrieval tasks. As such, their performance in this study should be seen as an exploration of their potential in a new context rather than a definitive assessment of their value to the field. The WSI-to-WSI matching methodology employed by the Yottixel search engine, which patches WSIs and applies a median-of-minimum strategy, may influence the observed performance. However, literature does not provide any other patch-based scheme for WSI-to-WSI comparison.

The challenge of representing an entire WSI with a single vector remains unresolved in computational pathology. The aggregation technique used in the GigaPath-WSI model shows some improvements, but we were not able to analyze all results. The results for GigaPath WSI have been provided for 17 organs and remaining organs are delayed due to the significant computational resources required for processing.

Hardware constraints currently necessitate the use of patch-based approaches and aggregation methods for WSI analysis. As technology advances and AI models grow in their capability to process larger inputs, it may become feasible to process entire WSIs in a single pass, potentially enabling more holistic representation learning approaches.

While the performance of foundation models in zero-shot WSI retrieval does not yet meet clinical standards, these models may indeed represent significant advancements in patch-level representation learning for histopathology. The challenges identified in this study present opportunities for future research, particularly in WSI-level representation learning, novel aggregation techniques, and retrieval methodologies optimized for histopathology. Significant refinements and alternative strategies are still needed to achieve reliable and accurate cancer tissue subtyping through image search and retrieval.

# References

[Bommasani2021] Bommasani, Rishi, Drew A. Hudson, Ehsan Adeli, Russ Altman, Simran Arora, Sydney von Arx, Michael S. Bernstein et al. "On the opportunities and risks of foundation models." arXiv preprint arXiv:2108.07258 (2021).

[Lahr2024] Lahr I, Alfasly S, Nejat P, Khan J, Kottom L, Kumbhar V, Alsaafin A, Shafique A, Hemati S, Alabtah G, Comfere N, Murphree D, Mangold A, Yasir S, Meroueh C, Boardman L, Shah VH, Garcia JJ, Tizhoosh HR. Analysis and Validation of Image Search Engines in Histopathology. IEEE Rev Biomed Eng. 2024 Jul 12;PP. doi: 10.1109/RBME.2024.3425769. Epub ahead of print. PMID: 38995713.

[Kalra2020a] Kalra, S., Tizhoosh, H. R., Choi, C., Shah, S., Diamandis, P., Campbell, C. J., & Pantanowitz, L. (2020). Yottixel–an image search engine for large archives of histopathology whole slide images. Medical Image Analysis, 65, 101757.

[Tizhoosh2024] Tizhoosh, H. R., & Pantanowitz, L. (2024). On image search in histopathology. Journal of Pathology Informatics, 100375.

[Kalra 2020b] Kalra, S., Tizhoosh, H. R., Shah, S., Choi, C., Damaskinos, S., Safarpoor, A., ... & Pantanowitz, L. (2020). Pan-cancer diagnostic consensus through searching archival histopathology images using artificial intelligence. NPJ digital medicine, 3(1), 31., Chicago

[Chen2024] Chen, R. J., Ding, T., Lu, M. Y., Williamson, D. F., Jaume, G., Song, A. H., ... & Mahmood, F. (2024). Towards a general-purpose foundation model for computational pathology. Nature Medicine, 30(3), 850-862.

[Vorontsov2024] Vorontsov, E., Bozkurt, A., Casson, A., Shaikovski, G., Zelechowski, M., Severson, K., ... & Fuchs, T. J. (2024). A foundation model for clinical-grade computational pathology and rare cancers detection. Nature Medicine, 1-12.

[Xu2024] Xu, H., Usuyama, N., Bagga, J., Zhang, S., Rao, R., Naumann, T., ... & Poon, H. (2024). A whole-slide foundation model for digital pathology from real-world data. Nature, 1-8.

# Appendix

### Table A1. Soft Tissues

| Primary Diagnosis | Patients | WSIs |
|---|---|---|
| Dedifferentiated liposarcoma | 57 | 87 |
| Fibromyxosarcoma | 24 | 141 |
| Leiomyosarcoma, NOS | 97 | 152 |
| Malignant fibrous histiocytoma | 12 | 54 |
| Malignant peripheral nerve sheath tumor | 9 | 27 |
| Synovial sarcoma, spindle cell | 6 | 12 |
| Undifferentiated sarcoma | 34 | 107 |

### Table A2. Rectum

| Adenocarcinoma, NOS | 133 | 134 |
|---|---|---|
| Adenocarcinoma in tubulovillous adenoma | 8 | 8 |
| Mucinous adenocarcinoma | 15 | 15 |
| Tubular adenocarcinoma | 5 | 5 |

### Table A3. Lining of the chest or abdomen

| Epithelioid mesothelioma, malignant | 51 | 62 |
|---|---|---|
| Mesothelioma, biphasic, malignant | 18 | 19 |
| Mesothelioma, malignant | 5 | 5 |

### Table A4. Thymus

| Thymic carcinoma, NOS | 11 | 11 |
|---|---|---|
| Thymoma, type A, malignant | 14 | 25 |
| Thymoma, type AB, NOS | 7 | 7 |
| Thymoma, type AB, malignant | 31 | 39 |
| Thymoma, type B1, malignant | 12 | 35 |
| Thymoma, type B2, NOS | 5 | 5 |
| Thymoma, type B2, malignant | 25 | 43 |
| Thymoma, type B3, malignant | 13 | 13 |

### Table A5. Pancreas

| | | |
|---|---|---|
| Adenocarcinoma, NOS | 18 | 18 |
| Infiltrating duct carcinoma, NOS | 150 | 176 |
| Mucinous adenocarcinoma | 5 | 5 |
| Neuroendocrine carcinoma, NOS | 8 | 8 |

### Table A6. Thyroid gland

| | | |
|---|---|---|
| Nonencapsulated sclerosing carcinoma | 4 | 5 |
| Papillary adenocarcinoma, NOS | 355 | 365 |
| Papillary carcinoma, columnar cell | 38 | 38 |
| Papillary carcinoma, follicular variant | 105 | 107 |

### Table A7. Cervix

| | | |
|---|---|---|
| Adenocarcinoma, NOS | 7 | 7 |
| Adenocarcinoma, endocervical type | 18 | 18 |
| Adenosquamous carcinoma | 5 | 6 |
| Mucinous adenocarcinoma, endocervical type | 17 | 17 |
| Squamous cell carcinoma, NOS | 157 | 163 |
| Squamous cell carcinoma, keratinizing, NOS | 24 | 27 |
| Squamous cell carcinoma, large cell, nonkeratinizing, NOS | 36 | 36 |

### Table A8. Prostate

| | | |
|---|---|---|
| Adenocarcinoma, NOS | 399 | 445 |
| Infiltrating duct carcinoma, NOS | 4 | 4 |

### Table A9. Eye

| | | |
|---|---|---|
| Epithelioid cell melanoma | 12 | 12 |
| Mixed epithelioid and spindle cell melanoma | 39 | 39 |
| Spindle cell melanoma, NOS | 19 | 19 |
| Spindle cell melanoma, type B | 9 | 9 |

### Table A10. Colon

| | | |
|---|---|---|
| Adenocarcinoma, NOS | 382 | 389 |

| Mucinous adenocarcinoma | 62 | 63 |

### Table A11. Testicles

| None | 16 | 43 |
|---|---|---|
| Embryonal carcinoma, NOS | 26 | 42 |
| Mixed germ cell tumor | 27 | 50 |
| Seminoma, NOS | 66 | 94 |
| Teratoma, benign | 5 | 13 |
| Yolk sac tumor | 4 | 6 |

### Table A12. Uterus

| Carcinosarcoma, NOS | 11 | 19 |
|---|---|---|
| Endometrioid adenocarcinoma, NOS | 376 | 416 |
| Mullerian mixed tumor | 45 | 71 |
| Papillary serous cystadenocarcinoma | 4 | 4 |
| Serous cystadenocarcinoma, NOS | 118 | 134 |

### Table A13. Esophagus

| Adenocarcinoma, NOS | 65 | 65 |
|---|---|---|
| Squamous cell carcinoma, NOS | 84 | 86 |
| Squamous cell carcinoma, keratinizing, NOS | 5 | 5 |

### Table A14. Bladder

| Papillary transitional cell carcinoma | 66 | 66 |
|---|---|---|
| Transitional cell carcinoma | 317 | 388 |

### Table A15. Adrenal gland

| Adrenal cortical carcinoma | 56 | 227 |
|---|---|---|
| Extra-adrenal paraganglioma, NOS | 17 | 18 |
| Extra-adrenal paraganglioma, malignant | 6 | 6 |
| Paraganglioma, NOS | 4 | 4 |
| Pheochromocytoma, NOS | 110 | 117 |
| Pheochromocytoma, malignant | 37 | 47 |

### Table A16. Liver

| Combined hepatocellular carcinoma and cholangiocarcinoma | 6 | 6 |
|---|---|---|
| Hepatocellular carcinoma, NOS | 350 | 364 |
| Hepatocellular carcinoma, clear cell type | 4 | 4 |

### Table A17. Skin

| Amelanotic melanoma | 7 | 7 |
|---|---|---|
| Epithelioid cell melanoma | 7 | 7 |
| Malignant melanoma, NOS | 384 | 424 |
| Nodular melanoma | 21 | 22 |
| Superficial spreading melanoma | 5 | 5 |

### Table A18. Lungs

| Acinar cell carcinoma | 19 | 19 |
|---|---|---|
| Adenocarcinoma, NOS | 293 | 334 |
| Adenocarcinoma with mixed subtypes | 98 | 104 |
| Basaloid squamous cell carcinoma | 13 | 13 |
| Bronchio-alveolar carcinoma, mucinous | 5 | 5 |
| Bronchiolo-alveolar carcinoma, non-mucinous | 16 | 16 |
| Mucinous adenocarcinoma | 12 | 16 |
| Papillary adenocarcinoma, NOS | 20 | 29 |
| Papillary squamous cell carcinoma | 4 | 4 |
| Solid carcinoma, NOS | 6 | 6 |
| Squamous cell carcinoma, NOS | 445 | 479 |
| Squamous cell carcinoma, keratinizing, NOS | 13 | 13 |

### Table A19. Stomach

| Adenocarcinoma, NOS | 150 | 153 |
|---|---|---|
| Adenocarcinoma, intestinal type | 82 | 84 |
| Carcinoma, diffuse type | 65 | 72 |
| Mucinous adenocarcinoma | 22 | 23 |
| Papillary adenocarcinoma, NOS | 8 | 9 |
| Signet ring cell carcinoma | 14 | 14 |
| Tubular adenocarcinoma | 73 | 85 |

Table A20. Head and Neck

| | | |
|---|---|---|
| Basaloid squamous cell carcinoma | 9 | 9 |
| Squamous cell carcinoma, NOS | 383 | 400 |
| Squamous cell carcinoma, keratinizing, NOS | 52 | 56 |
| Squamous cell carcinoma, large cell, nonkeratinizing, NOS | 5 | 6 |

Table A21. Breast

| | | |
|---|---|---|
| Infiltrating duct and lobular carcinoma | 26 | 27 |
| Infiltrating duct carcinoma, NOS | 757 | 805 |
| Infiltrating duct mixed with other types of carcinoma | 19 | 20 |
| Infiltrating lobular mixed with other types of carcinoma | 7 | 7 |
| Intraductal papillary adenocarcinoma with invasion | 6 | 7 |
| Lobular carcinoma, NOS | 191 | 204 |
| Medullary carcinoma, NOS | 6 | 7 |
| Metaplastic carcinoma, NOS | 13 | 13 |
| Mucinous adenocarcinoma | 16 | 20 |

Table A22. Brain

| | | |
|---|---|---|
| Astrocytoma, NOS | 59 | 104 |
| Astrocytoma, anaplastic | 122 | 164 |
| Glioblastoma | 389 | 860 |
| Mixed glioma | 130 | 216 |
| Oligodendroglioma, NOS | 107 | 204 |
| Oligodendroglioma, anaplastic | 72 | 155 |

Table A23. Kidneys

| | | |
|---|---|---|
| Clear cell adenocarcinoma, NOS | 500 | 506 |
| Papillary adenocarcinoma, NOS | 276 | 300 |
| Renal cell carcinoma, NOS | 13 | 13 |
| Renal cell carcinoma, chromophobe type | 109 | 121 |